\documentclass{PoS}
\newcommand{\whizard}{\texttt{WHIZARD}}

\title{QCD NLO with Powheg matching and top threshold matching in
  WHIZARD} 

\ShortTitle{QCD NLO in WHIZARD}

\author{\speaker{J\"urgen Reuter}\\
        DESY Theory Group, Notkestr. 85, 22607 Hamburg, Germany\\
        E-mail: \email{juergen.reuter@desy.de}}

\author{Fabian Bach\\
        European Commission, Eurostat, 2920 Luxembourg, Luxembourg\\
        E-mail: \email{fabian.bach@t-online.de}}

\author{Bijan Chokouf\'{e} Nejad\\
        DESY Theory Group, Notkestr. 85, 22607 Hamburg, Germany\\
        E-mail: \email{bijan.chokoufe@desy.de}}

\author{Wolfgang Kilian\\
        University of Siegen, Emmy-Noether-Campus, Walter-Flex-Str. 3,
        57068 Siegen, Germany\\ 
        E-mail: \email{kilian@physik.uni-siegen.de}}

\author{Maximilian Stahlhofen\\
        PRISMA Cluster of Excellence, Institute of Physics, 
        Johannes Gutenberg University, Staudingerweg 7, 55128 Mainz, 
        Germany \& DESY Theory Group, Notkestr. 85, 22607
        Hamburg, Germany \\
        E-mail: \email{mastahlh@uni-mainz.de}}

\author{Christian Weiss\\
        DESY Theory Group, Notkestr. 85, 22607 Hamburg, Germany \& 
        University of Siegen, Emmy-Noether-Campus, Walter-Flex-Str. 3,
        57068 Siegen, Germany\\
        E-mail: \email{christian.weiss@desy.de}}

\abstract{We present the status of the automation of NLO processes
  within the event generator WHIZARD. The program provides an
  automated FKS subtraction and phase space integration over the FKS
  regions, while the (QCD) NLO matrix element is accessed via the
  Binoth Les Houches Interface from an externally linked one-loop
  program. Massless and massive test cases and validation are shown
  for several $e^+e^-$ processes. Furthermore, we discuss work in
  progress and future plans. The second part covers the
  matching of the NRQCD prediction with NLL threshold resummation to
  the NLO continuum top pair production at lepton colliders. Both the
  S-wave and P-wave 
  production of the top pair are taken into account in the resummation. 
  The inclusion in
  WHIZARD allows to study more exclusive observables than just the
  total cross section and automatically accounts for important
  electroweak and relativistic corrections in the threshold region. 
 \begin{flushright}
    \normalsize{} DESY 15--176, MITP/16-007, SI-HEP-2016-02
 \end{flushright}}

\FullConference{12th International Symposium on Radiative Corrections
  (Radcor 2015) and LoopFest XIV (Radiative Corrections for the LHC
  and Future Colliders) \\ 15-19 June, 2015 \\ UCLA Department of
  Physics \& Astronomy Los Angeles, USA }

\begin{document}

\section{Introduction}

Monte Carlo event generators are indispensable tools for the
comparison of data from high energy physics experiments with theory
predictions. They should reflect the highest possible degree of
precision from calculations in perturbative quantum field theory. In
this proceedings article we describe several facets of the inclusion
and automation of such higher-order calculations into the code
\whizard. \whizard~\cite{Kilian:2007gr} is a multi-purpose event
generator, that comes with its own powerful (tree-level) matrix
element generator \texttt{O'Mega}~\cite{Moretti:2001zz}.
\texttt{O'Mega} generates matrix elements for arbitrary processes 
either as Fortran 95/2003 code that is compiled or as byte code
commands that are interpreted by a virtual machine
(OVM)~\cite{Nejad:2014sqa}. Integration is performed
using a multi-channel adaptive Monte-Carlo integration using its
subpackage VAMP~\cite{Ohl:1998jn}. As \whizard\ has been used for a
long time as a major workhorse for lepton collider simulations, it
contains the special tool \texttt{CIRCE1/2}~\cite{Ohl:1996fi} that
allows for the simulation of lepton collider beamstrahlung. 

Especially at hadron colliders like the Large Hadron Collider (LHC),
but also at high-energy lepton colliders, a precise treatment of QCD
effects is indispensable. Color information is treated in \whizard\
using the color flow formalism~\cite{Kilian:2012pz}. \whizard\ comes
with its own QCD parton showers, a $k_T$-ordered shower and an
analytic parton shower~\cite{Kilian:2011ka}. \whizard\ ships with the
final Fortran version of the \texttt{Pythia6}
package~\cite{Sjostrand:2006za} for hadronization, whose parton
showers can be used alternatively to the internal ones. Direct
interfaces to the standard event formats (like LHE, HepMC, LCIO,
StdHEP) do exist, as well as interfaces for jet clustering and parton
distribution functions. The most prominent PDFs are directly included
without the need to take them from an external library.
\whizard\ has a big tradition for studies of beyond the Standard Model
physics
(e.g.~\cite{Kilian:2004pp,Hagiwara:2005wg,Beyer:2006hx,Alboteanu:2008my,Kalinowski:2008fk,Kilian:2014zja,Kilian:2015opv,Kim:2015ron}),
we will here concentrate on the (automated) inclusion of higher order
(QCD) corrections to SM processes. 


\section{Automation of QCD NLO corrections}

In order to be able to make reliable quantitative predictions for any
kind of process, one has to include higher orders in the perturbative
series. For decades, this has been (and still is) a tedious job, but
in order to facilitate the procedure and to get to final results
quicker, a high degree of standardization and even automation is
highly desirable. In order to get a finite result that can be treated
in a Monte Carlo integration, the soft and collinear (infrared)
singularities as well as the ultraviolet divergences have to be
treated before getting into phase space integration. The latter is
done by means of renormalization in the virtual matrix elements. The
former ones cancel in final states that are inclusive enough due to
the KLN theorem~\cite{Kinoshita:1962ur,Lee:1964is}. However, for this
to happen, Born processes and virtual corrections have to be combined
with real radiation corrections. In order to make the pieces that
depend on final states of different multiplicites independently
finite, so called {\em subtraction formalisms} have been
invented. Here, the soft and collinear divergent regions 
are subtracted from the real emission diagrams to yield a finite
result, while the same subtraction terms integrated analytically over
the phase space of the emitted parton is added to the sum of Born and
virtual contributions to make them finite as well. After this, all
parts are finite and can be treated within the Monte Carlo
integration.  

The first attempt to include (virtual) next-to-leading order (NLO)
corrections in \whizard\ was in the context of QED corrections to
production of charginos at the ILC~\cite{Kilian:2006cj,Robens:2008sa}
not using a subtraction formalism, but indeed a phase-space slicing plus
photon resummation to obtain positive-weight NLO events. Virtual QCD
corrections have for the first time been included in \whizard\ using
Catani-Seymour subtraction~\cite{Catani:1996vz}
in~\cite{Binoth:2009rv,Greiner:2011mp} for the process $pp \to b\bar b
b \bar b + X$. This was using external virtual matrix elements from
(the predecessors of) the \texttt{Gosam}
package~\cite{Cullen:2014yla}, and a hard-coded subtraction
tailor-made for this specific process. 
Since version 2.2 of \whizard\, a preliminary version of the automated
implementation of FKS matching~\cite{Frixione:1995ms,Frixione:2009yq}
has been published. According to the FKS prescription, for all
\begin{figure}[htbp]
\centering
\includegraphics[width=0.45\textwidth]{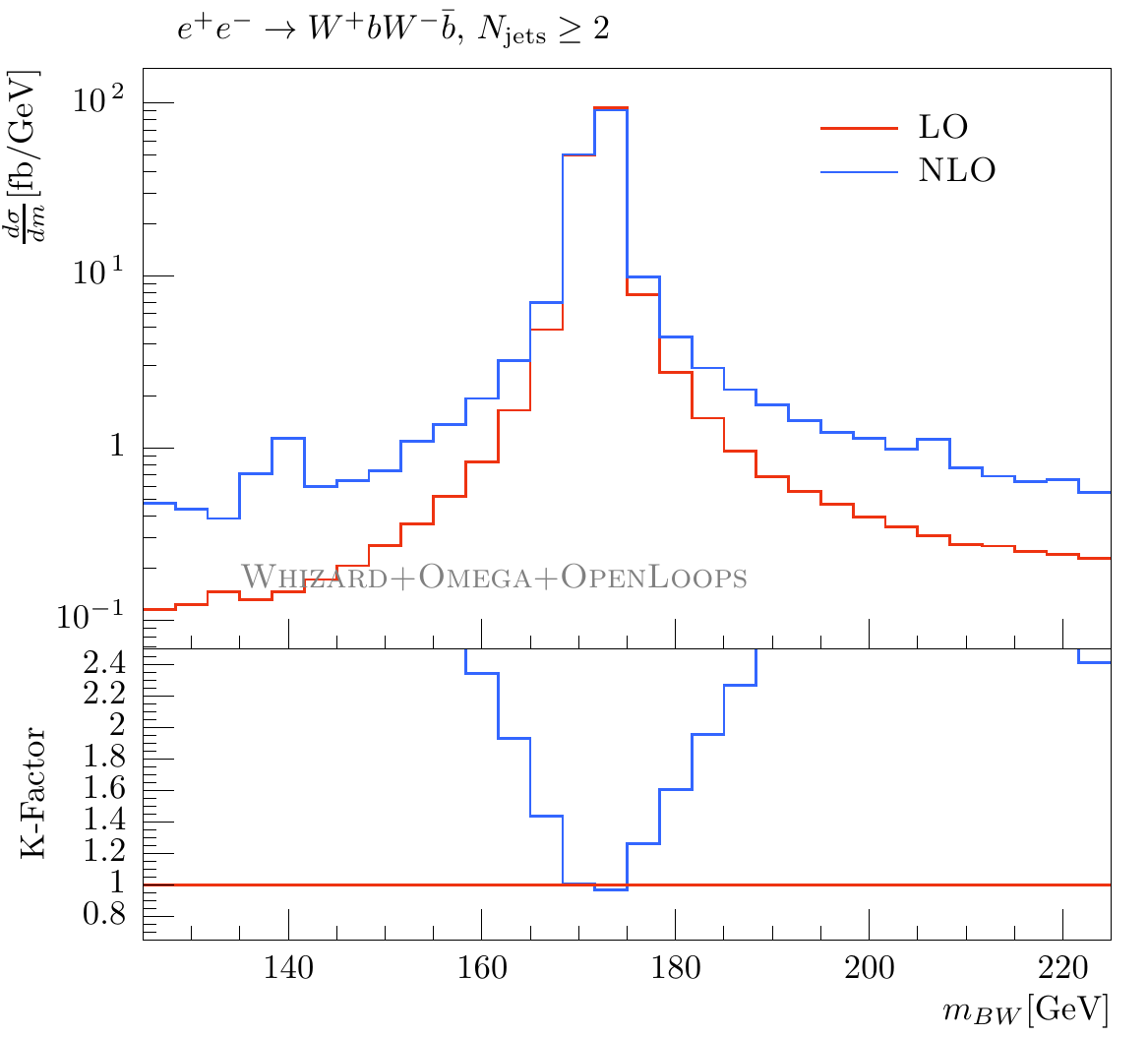}
\includegraphics[width=0.45\textwidth]{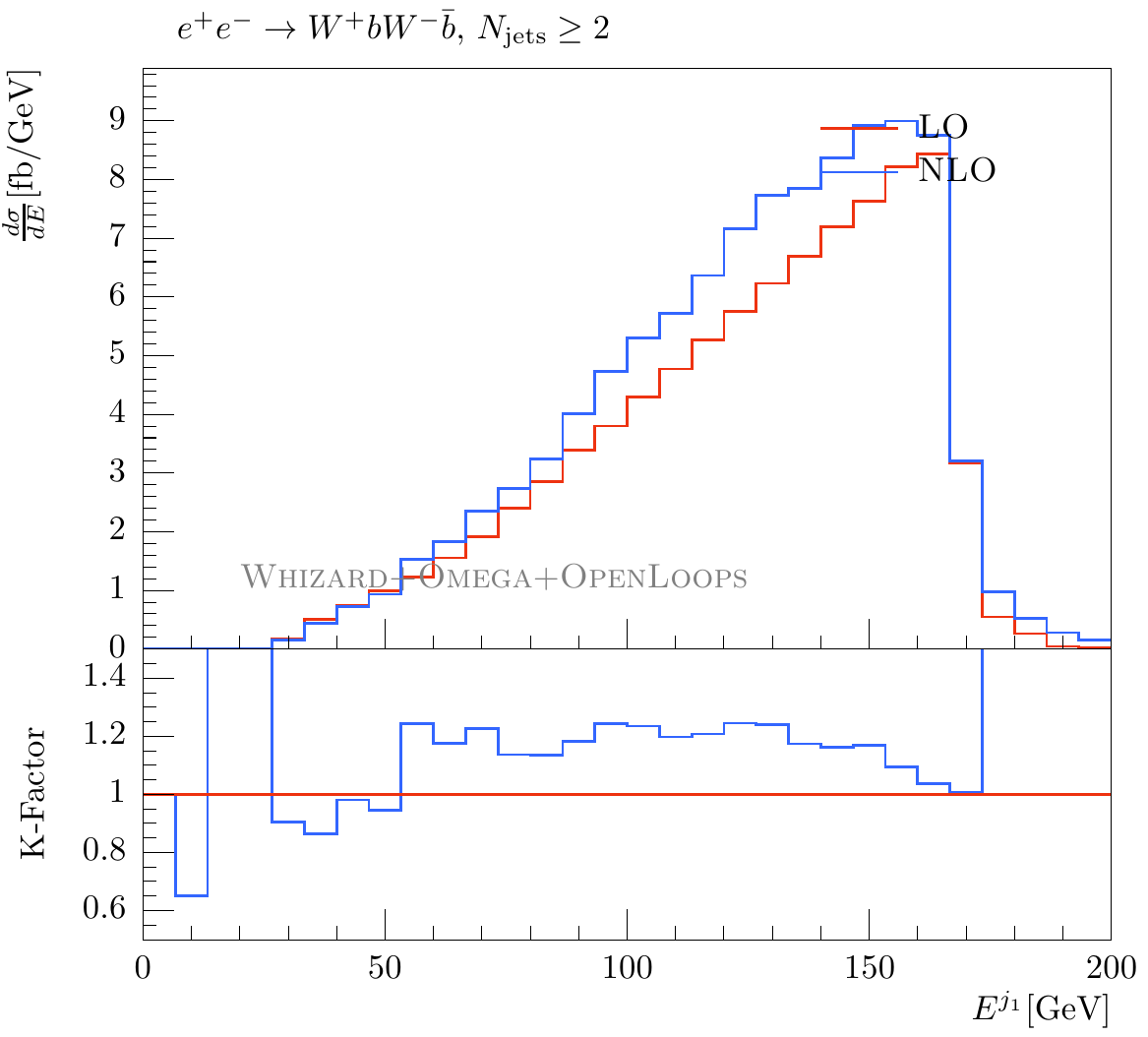}
\caption{Resonant and non-resonant top pair production, $e^+e^- \to
  W^+ b W^- \bar b$ at a future ILC at $\sqrt{s} = 500$ GeV. The left
  plot shows the $Wb$ invariant mass distribution (where $b$ is the
  jet that contains the $b$ quark), while the plot on
  the right-hand side shows the energy of the hardest jet. Red lines
  are LO, blue ones NLO, respectively.}
\label{fig:wbwb}
\end{figure}
possible singular regions, subtraction terms with their corresponding
phase space mappings are being generated. It automatically takes into
account the color-correlated matrix elements for soft subtractions,
and spin-correlated matrix elements for the collinear
subtractions. External virtual matrix elements can be taken either
from \texttt{Gosam} or \texttt{OpenLoops}~\cite{Cascioli:2011va}
interfaced via the Binoth Les Houch Accord
(BLHA)~\cite{Binoth:2010xt,Alioli:2013nda}. The settings for NLO
processes can be directly steered via commands within the SINDARIN
command language of \whizard. A BLHA contract file to 
external one-loop providers is then automatically created by
\whizard. Matrix elements from \texttt{Gosam} and \texttt{OpenLoops}
have been checked against each other. Many different processes with
uncolored initial states 
have been validated, like $e^+e^- \to q\bar q$, $e^+e^- \to q\bar q
g$, $e^+e^- \to \ell^+\ell^-q\bar q$, $e^+e^- \to \ell^+\nu_\ell q\bar
q$, $e^+e^- \to t\bar t$, $e^+e^- \to tW^- b$, $e^+e^- \to W^+ bW^-
\bar b$, $e^+e^- \to t\bar t H$, and $e^+ e^- \to W^+ b W^- \bar b
H$. Cuts can be applied at the level of the clustered objects for the
NLO processes. Fig.~\ref{fig:wbwb} shows as prime example differential
distributions for the resonant and non-resonant top pair production
for a 500 GeV ILC. Virtual matrix elements have been taken from 
\texttt{OpenLoops}. The plot on the left-hand side shows the $Wb$
invariant mass distribution which clearly shows the interference with
the non-resonant background at NLO (this depends, of course, on the
parameters of the jet clustering algorithm). The right hand side shows
the energy distribution of the hardest jet in the final state. 
As the ILC will always run with beam polarization~\cite{Baer:2013cma},
\whizard\ can handle polarized cross sections at the LO, but also at
the NLO level. Corresponding demands for polarized virtual matrix
elements have to be stated in the BLHA contract file. This is beyond
the standard BLHA convention and at the moment can be only processed
with \texttt{OpenLoops}. It is also
\begin{figure}[htbp]
\centering
\includegraphics[width=0.45\textwidth]{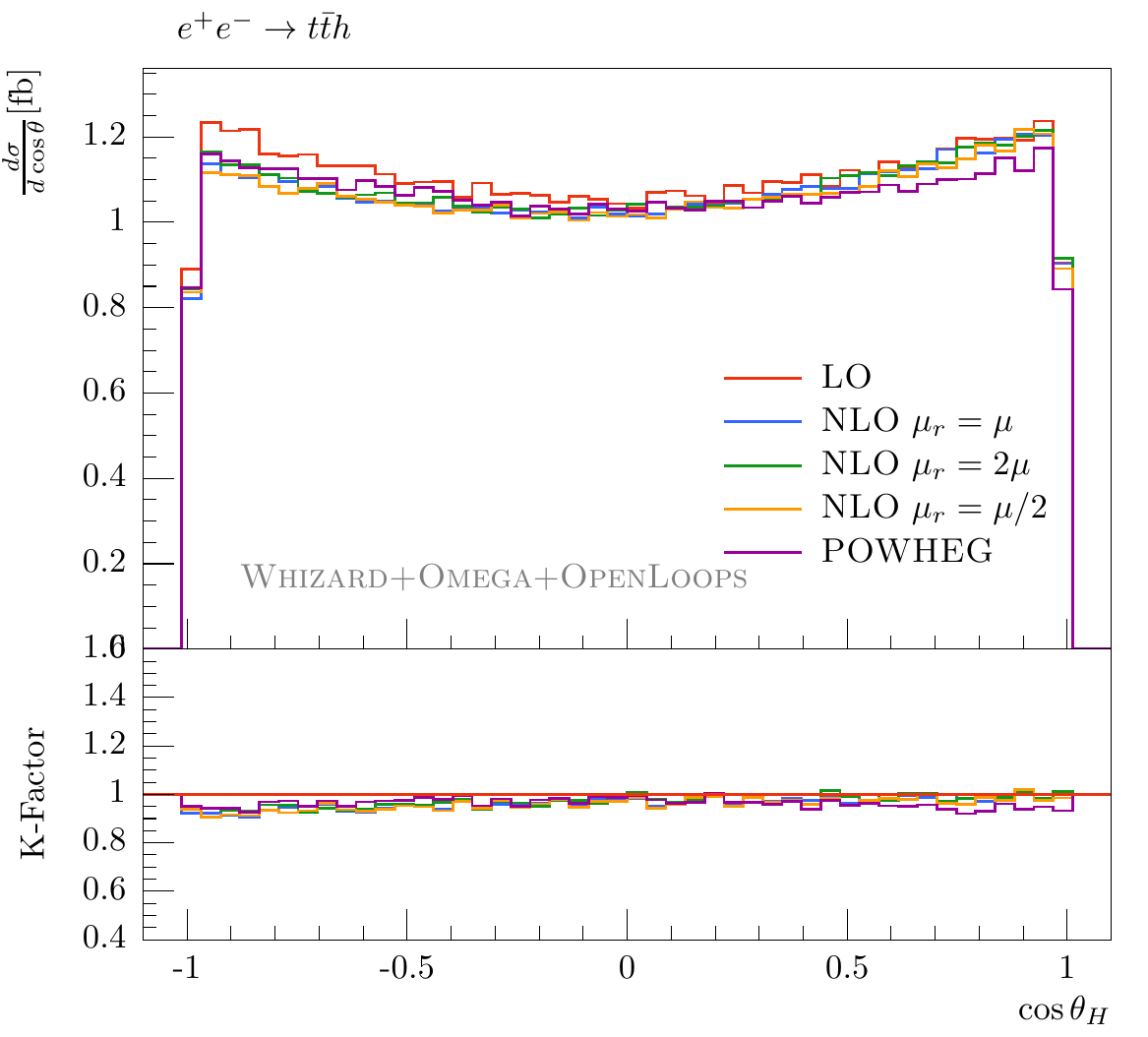}
\includegraphics[width=0.45\textwidth]{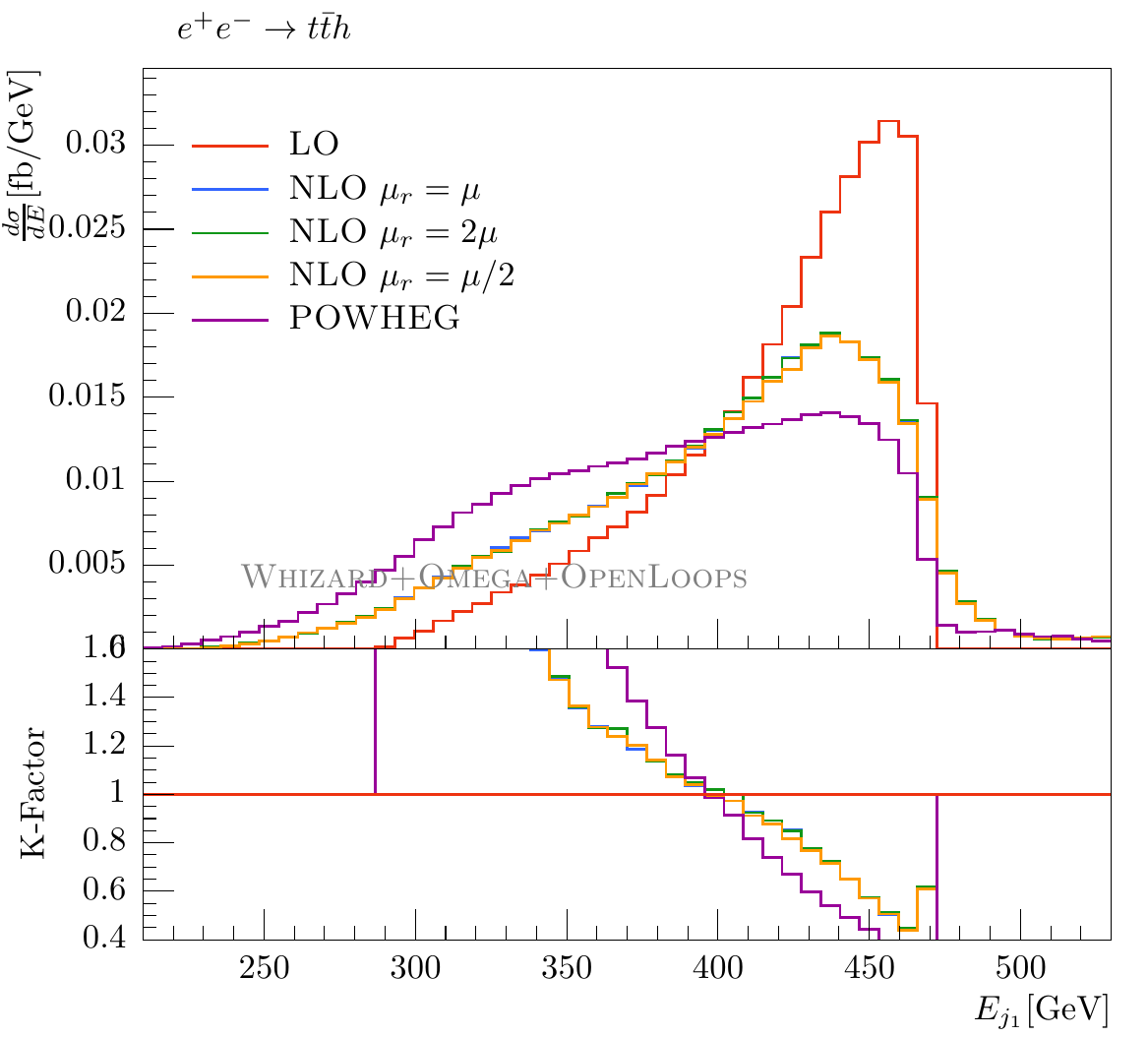}
\caption{Angular distribution of the Higgs boson and energy
  distribution of the leading jet in the process $e^+e^- \to t \bar t
  H$ at $\sqrt{s} = 1000$ GeV. The dark magenta curve shows the POWHEG
  matching.}
\label{fig:tth}
\end{figure}
easily possible to integrate NLO processes that are not happening in
the lab system due to initial state (photon) radiation or
beamstrahlung. Also, decay processes can be integrated and simulated
at the NLO-level, where the automated subtraction has to take care to 
generated subtraction terms and phase space in such a way as to
maintain the mass of the decaying particle. This is important to have
widths for resonant NLO processes in the same scheme/setup, as e.g. in
the top threshold simulation in the next section.
In the case that matrix elements contain narrow resonances decaying
into colored objects, e.g. $Z\to q\bar q$ or $H\to b\bar b$, then soft
and collinear radiation at NLO can drag final state particles either
away from a resonance or push them towards a resonance. Both cases
leads to a mismatch within the subtraction terms and a usually awful
convergence of the Monte Carlo integration. Ref.~\cite{Jezo:2015aia}
proposed a method to remedy this problem which has been applied to
a specific process and by using a hard-coded resonance history in that
reference. \whizard\ now has an automated implementation of this
algorithm that generates the necessary resonance history. This will be
included in the upcoming version \texttt{WHIZARD 2.3}. 
This formalism is particularly useful for the application of matrix
element-parton shower matching methods at NLO, like e.g. POWHEG
matching~\cite{Nason:2004rx}. An automatic implementation of POWHEG
matching is implemented in \whizard\ that combines (matches) the NLO
fixed-order matrix element with the parton shower in such a way that
the hardest emission is correctly described by the NLO matrix
element. In Fig.~\ref{fig:tth} we show the process $e^+e^-\to t\bar t H$
for $\sqrt{s} = 1000$ GeV at LO, at NLO using different scale choices
and the POWHEG-matched result. As expected, the more inclusive angular
distribution of the Higgs boson is rather insensitive to the QCD
corrections, while the energy distribution of the leading jet depends
strongly on it. Particularly, the Sudakov suppression at high energy
values can be observed.

Before work in progress and future plans on the automation of
fixed order NLO will be discussed in the summary,
Sec.~\ref{sec:conclusions}, we describe the implementation and
matching of threshold resummation for the lepton collider top
threshold and the continuum fixed-order NLO calculation.


\section{Top Threshold Matching at Lepton Colliders}

High-energy lepton colliders that can be operated at or above the top
threshold allow for the most precise method to measure the mass of the
top quark known to date.
In particular they allow for a direct extraction of the mass in a 
appropriate (short distance) scheme that is free of QCD renormalon 
uncertainties. 
For this, a threshold scan is performed as relatively
inclusive measurement, and from the fit of the cross section line-shape 
to data the top mass can be extracted with an uncertainty well below 
100 MeV. On the other hand, due to cuts and
tagging in the final state, the measurement might not be as inclusive
as anticipated, and differential quantities might allow for further
improvements of the measurements at the top threshold.
In particular this may be relevant for the precise measurement of other 
important (SM) parameters like the top width (i.e. $V_{tb}$), the strong 
coupling $\alpha_s$ or the top Yukawa coupling.
Hence, it is desirable to have a Monte Carlo prediction for the threshold 
region.
Furthermore, future lepton colliders might run at energies
close to, but a bit off threshold (cf. e.g. the 380 GeV staging from
the CLIC study group). 
Therefore a smooth transition (matching) between the resummed threshold 
prediction and the fixed order prediction at large energies is required.

The top quark decays through the electroweak
interaction before it hadronizes, but at the top threshold the
attractive top quark potential leaves a remnant of a (1S) toponium bound
state. 
The bound state effects are reflected by Coulomb singular terms 
$\propto (\alpha_s/v)^n$ in the perturbative expansion of the cross 
section, where $v \sim \alpha_s \sim 0.1$ is the relative velocity of the 
top quarks close to threshold.
In addition to the Coulomb singularities, large logarithms 
$\propto \ln^n v$ invalidate conventional fixed-order perturbation theory 
in the threshold region.
These singular terms can be systematically resummed in
vNRQCD~\cite{Luke:1999kz}, an effective field theory where modes at
the hard scale, $m_t$, are integrated out. (Further scales in the
theory are the soft scale, given by the the top momentum, $m_t v$, and
the ultrasoft scale, $m_t v^2$, given by the kinetic energies of the
tops).
\begin{figure}[htbp]
\centering
\includegraphics[width=0.95\textwidth]{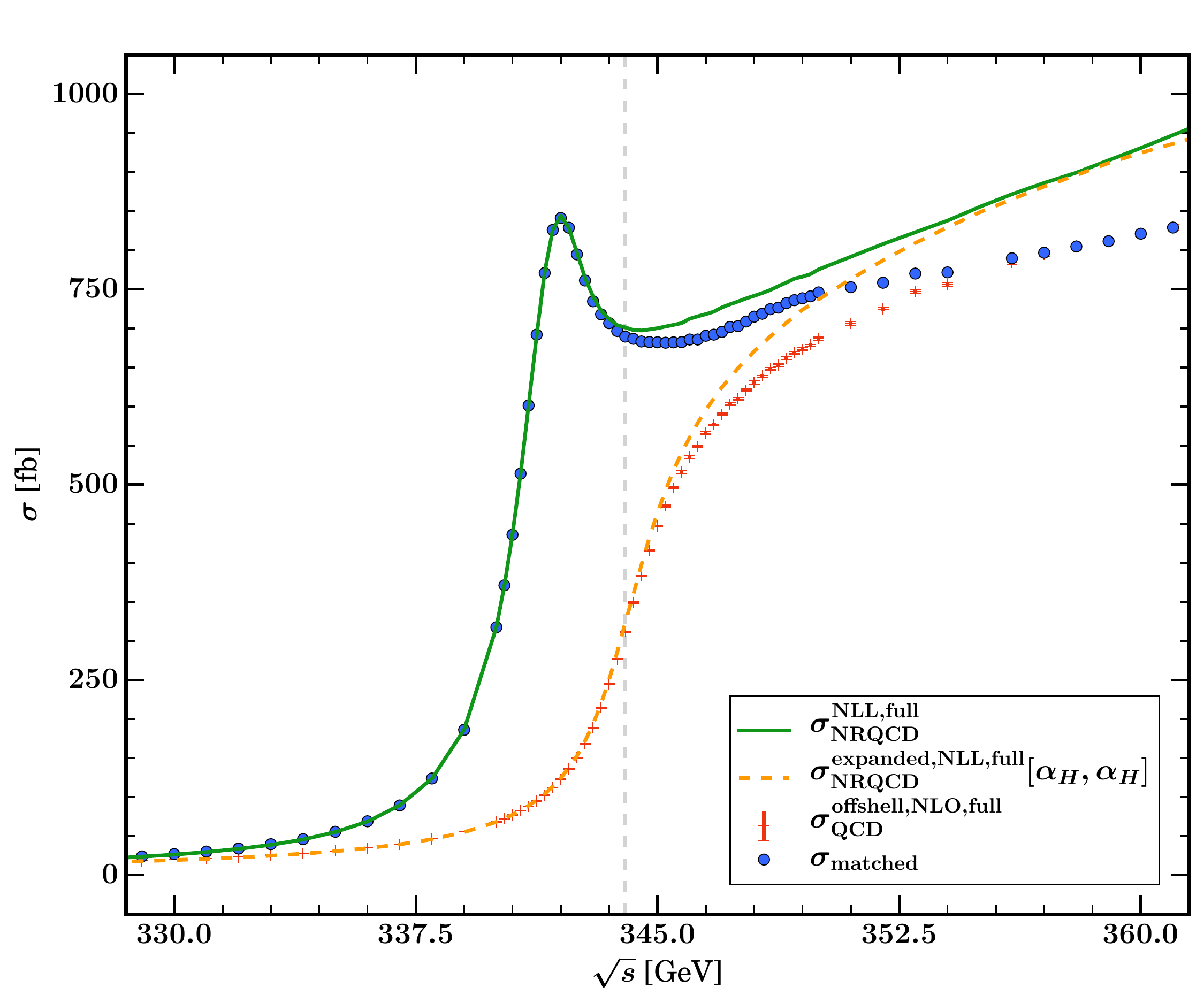}
\caption{Matching of the NLL resummed threshold prediction to the
  fixed-order NLO QCD continuum for the total cross section of the
  process $e^+e^- \to W^+ b W^- \bar b$. The red points are the
  NLO cross section values, the green
  curve corresponds to the inclusion of the NLL threshold resummed form 
  factor into the Born process. 
  The dashed orange curve corresponds to the latter result with
  the form factor expanded to first order in $\alpha_s$ and evaluated at
  the hard scale. (The two values of 
  $\alpha_s$ are used in the hard matching coefficient and the Coulomb 
  Green function of the form factor, respectively.)
  The blue dots show the result of our (preliminary) matching procedure.
  For this plot we have chosen $m_t= 172$ GeV as well as the other
  parameters of Table 1 in~\cite{Bach:2014hxa}.}
\label{fig:tt_threshold}
\end{figure}

We implement the vNRQCD resummation in the Monte Carlo process by 
replacing the SM $tt\gamma, ttZ$ vertices by corresponding (S-wave 
and P-wave) non-relativistic form factors at next-to-leading
logarithmic (NLL) order using the TOPPIK code~\cite{Hoang:1999zc}. 
At the same time we include the full set 
of relativistic QCD corrections at NLO, respectively.
That is to say, we consistently add the terms beyond NLO from the 
NLL threshold resummation to the fixed order result.
For technical details of the (v)NRQCD calculations, 
cf.~\cite{Pineda:2001et,Hoang:2006ty,Hoang:2010gu,Hoang:2011gy,Hoang:2013uda}. 
The matching between the threshold and the (relativistic) NLO 
prediction is performed by switching off the resummation above the 
threshold and expanding the resummed result.
For onshell top quarks it is quite clear how to perform this as the
complete amplitude can be simply multiplied with the form factor.
For the non-resonant contributions in the process 
$e^+e^- \to W^+ b W^- \bar b$ it is quite nontrivial due to the 
interference of threshold-enhanced terms with the non-resonant background.
To maintain electroweak gauge invariance in the combination of the
form factor and the QCD NLO corrections to the decay, we factorize
production and decay in this contribution (similar to the pole
approximation). 

As consistency check we found agreement between the numerical \whizard\ 
implementation and the analytic non-relativistic vNRQCD result in the
limit $\alpha_s \to 0$ for small $v$. 
Furthermore, in the threshold region, the Born prediction with an insertion 
of the form factor expanded to $\mathcal{O}(\alpha_s)$ should coincide with the 
fixed-order NLO cross section, because the vNRQCD form factor contains 
the dominant (non-vanishing) terms as $v\to 0$.
This can be observed by the nice agreement of the dashed orange curve
and the red dots in Fig.~\ref{fig:tt_threshold} for $\sqrt{s}< 345\,
{\rm GeV}$. The green curve shows the prediction corresponding to the
insertion of the full NLL resummed form factor into the Born process.
As the blue dots we show the final cross section with a matching
prescription to interpolate between the non-relativistic threshold
region and the continuum. The resummation in the form factor has to be
switched off at large energies where the expansion in $v$ is no longer
valid.


\section{Summary and Outlook}
\label{sec:conclusions}

This article gives a status report on the automation of NLO QCD
corrections within the event generator \whizard. While virtual matrix
elements are taken from external libraries or programs, subtraction
terms are automatically generated for arbitrary SM processes according
to the FKS algorithm, together with the corresponding partitioning of
the phase space. Several cross checks, validations and examples have
been given for uncolored initial states. The subtraction terms for
initial-initial and initial-final singular regions are implemented and
validated. Structure function reweighting for parton distribution
functions in the initial state are up and working. Hence, all
ingredients for automated NLO QCD processes for hadron colliders are
present, but there have not been extensive enough checks and
validations as of now. 

Further steps in the more mid-term future will be the inclusion of QED
and electroweak automated NLO corrections. Besides the POWHEG matching
shown here, also merging schemes are being implemented in
\whizard. The ultimate goal here is to have automated merging
prescriptions for inclusive multi-jet samples. 

A related, but conceptually different topic is the matching between
the fixed-order QCD NLO matrix element for the continuum (off-shell)
top quark pair production at lepton colliders to the NLL threshold
production in vNRQCD. All technical and field-theoretical ingredients
for this process are understood and being implemented. The implementation
then allows for the simulation of more exclusive observables, and
directly accounts for non-leading electroweak corrections within the
implemented matrix elements. With our (preliminary) matching
procedure the transition of the total cross section from the threshold 
to the continuum region looks satisfactory. 

We plan to apply the same formalism for the $ttH$
production (off-shell) at the threshold, which this time is not driven
by physical arguments but by the design energy of 500 GeV of the
International Linear Collider.

\acknowledgments
We would like to thank E.~Bagnaschi, A.~Denner, N.~Greiner, A.~Hoang,
J.~Lindert, and S.~Pozzorini for their contribution to the projects
summarized in this proceedings article, for providing and helping us
with different codes like e.g. \texttt{Gosam} and \texttt{Openloops},
as well as for enlightening discussions on technical and physical
details within perturbative quantum field theory. JRR wants to thank
the organizers of Radcor/Loopfest 2015 for a great conference in the
thrilling summer of Southern California.

\end{document}